\documentclass[prl,aps,twocolumn]{revtex4}
\usepackage{graphicx}
\usepackage{epstopdf}
\usepackage{dcolumn}
\usepackage{bm}
\usepackage{color}
\usepackage{ulem}
\usepackage{amsfonts,amsthm}
\usepackage{amsmath}
\usepackage{amssymb}
\begin{document}
\newcommand{\bR}{\mbox{\boldmath $R$}}
\newcommand{\tr}[1]{\textcolor{black}{#1}}
\newcommand{\tb}[1]{\textcolor{black}{#1}}
\newcommand{\tc}[1]{\textcolor{black}{#1}}
\newcommand{\Ha}{\mathcal{H}}
\newcommand{\mh}{\mathsf{h}}
\newcommand{\mA}{\mathsf{A}}
\newcommand{\mB}{\mathsf{B}}
\newcommand{\mC}{\mathsf{C}}
\newcommand{\mS}{\mathsf{S}}
\newcommand{\mU}{\mathsf{U}}
\newcommand{\mX}{\mathsf{X}}
\newcommand{\sP}{\mathcal{P}}
\newcommand{\sL}{\mathcal{L}}
\newcommand{\sO}{\mathcal{O}}
\newcommand{\la}{\langle}
\newcommand{\ra}{\rangle}
\newcommand{\ga}{\alpha}
\newcommand{\gb}{\beta}
\newcommand{\gc}{\gamma}
\newcommand{\gs}{\sigma}
\newcommand{\vk}{{\bm{k}}}
\newcommand{\vq}{{\bm{q}}}
\newcommand{\vR}{{\bm{R}}}
\newcommand{\vQ}{{\bm{Q}}}
\newcommand{\vga}{{\bm{\alpha}}}
\newcommand{\vgc}{{\bm{\gamma}}}
\newcommand{\mb}[1]{\mathbf{#1}}
\def\vec#1{\boldsymbol #1}
\arraycolsep=0.0em
\newcommand{\Ns}{N_{\text{s}}}
%

\title{
{Finite-Temperature Signatures of Spin Liquids in Frustrated Hubbard Model}
}

\author{
Takahiro Misawa$^1$  and Youhei  Yamaji$^{2,3}$
}

\affiliation{$^1$Institute for Solid State Physics, The University of Tokyo, 5-1-5 Kashiwanoha, Kashiwa, Chiba 277-8581, Japan}
\affiliation{$^2$Quantum-Phase Electronics Center (QPEC), The University of Tokyo, Hongo, Bunkyo-ku, Tokyo, 113-8656, Japan}
\affiliation{$^3$JST, PRESTO, Hongo, Bunkyo-ku, Tokyo, 113-8656, Japan}

\date{\today}

\begin{abstract}
Finite-temperature properties of the frustrated Hubbard model 
are theoretically examined by
using the recently proposed thermal pure quantum state,
which is an unbiased numerical method
for finite-temperature calculations.
By performing systematic calculations for the frustrated
Hubbard model, we show that the 
geometrical frustration controls the characteristic energy scale
of the metal-insulator transitions.
{We also find that entropy remains large even at 
moderately high temperature around the region 
where the quantum spin liquid is expected to appear at zero temperature.}
We propose that this is a useful criterion whether
the target systems have a chance to be 
the quantum spin liquid or the non-magnetic insulator 
at zero temperature.
\end{abstract}


\maketitle


{{\it Introduction.}--}
Strong correlations among particles often induce
localization of the particles and resultant
{charge-gapped} states are called Mott insulators.
{The Mott insulators have been ubiquitously found 
in a broad range of condensed matter physics~\cite{ImadaRMP,Kanoda,BlochRMP}.}
In most of the Mott insulators in solids, 
{time-reversal symmetry-broken phases such as}
antiferromagnetic phases appear at {sufficiently} low temperatures.
{However, 
if geometrical frustration becomes large~\cite{Diep},}
{the} quantum melting of the magnetic orders leads to new states of matter
such as quantum spin liquids (QSL)~\cite{Shannon_PRL2012,Normand_PRL2014}.
Actually, in the several organic conductors,
it has been pointed out that QSL appear
~\cite{Shimizu2003,Itou2008,kanoda2011mott}.
It has been one of the hottest issues
of the modern condensed matter physics to clarify
how the interplay of strong electronic correlations
and the geometrical frustrations induces the 
QSL~{\cite{balents2010spin}}.

\begin{figure}[tb!]
  \begin{center}
    \includegraphics[width=7.5cm,clip]{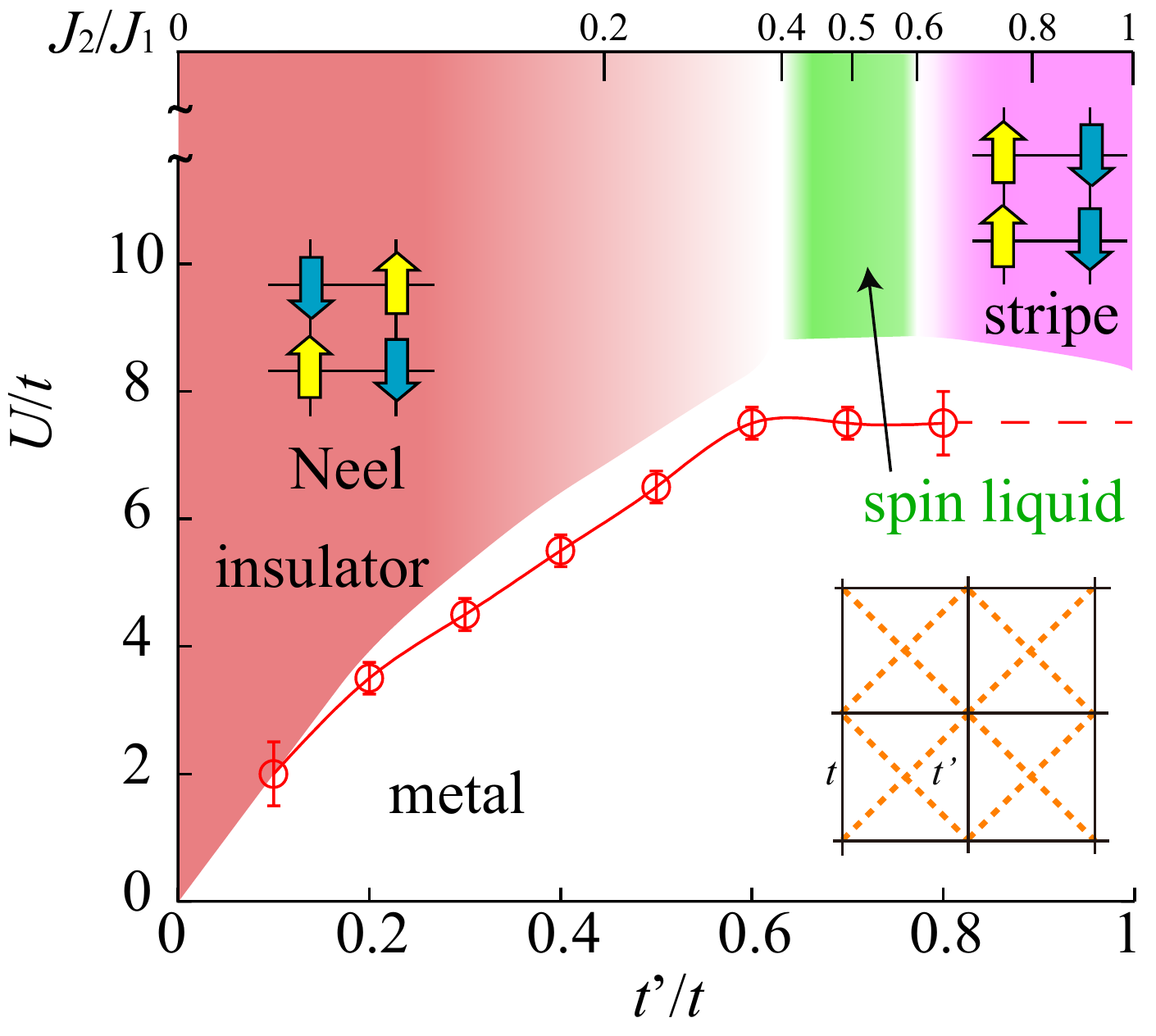}
  \end{center}
\caption{(color online).~Phase diagram for frustrated Hubbard model
in comparison with strong-coupling-limit phase diagram. The phase boundary separating 
insulating and metallic ground states
is determined by maxima of $\chi_{D}$ (see main article) as a function of on-site Coulomb repulsion $U$.
Spin liquid may appear between {N\'eel} and 
stripe magnetic orders {in the strong coupling region}.
In the inset, lattice structure used in this study is shown. 
The nearest-neighbor hopping (next-nearest-neighbor)
hopping is represented by $t$ ($t^{\prime}$).
}
\label{fig:schem}
\end{figure}

The two-dimensional Hubbard model with geometrical 
frustrations, 
{ which has the nearest-neighbor [nn]
(next-nearest-neighbor [nnn])} hopping $t$ ($t^{\prime}$) 
 and on-site Coulomb interaction $U$ 
(details are defined in Eq.~\eqref{eq:H} later)
is one of the simplest theoretical model that 
describes 
{interplay between}
the strong electronic
correlations and the geometrical frustrations.
In this model,
{due to $t^{\prime}$,}
which induces the {nnn} antiferromagnetic interactions, 
as illustrated 
in the inset of Fig.~\ref{fig:schem},
the competition between two magnetic phases occurs:~
a simple N\'eel state becomes stable for small $t^{\prime}$ while
a stripe state becomes stable for large $t^{\prime}$ 
region ($t^{\prime}/t\sim 1$).
{Several theoretical calculations for the ground states
of the frustrated Hubbard model~\cite{kashima2001magnetic,MizusakiImada,Tocchio2008} 
and its strong coupling limit 
$J_{1}$-$J_{2}$ Heisenberg model ($J_{1}\sim 4t^2/U, J_{2}\sim 4{t^{\prime}}^2/U$)~\cite{Jiang2012,Hu2013,Gong2014,Morita2015}
have been done {thus} far
and most of {the} calculations suggest that QSL {states} 
{appear} around the intermediate region. 
In spite of the huge amount of the studies on the 
frustrated Hubbard model {and} $J_{1}$-$J_{2}$ Heisenberg model,
there are few unbiased theoretical
studies on the finite-temperature properties
{that are accessible in experiments}
{because of a} lack of efficient theoretical {methods}.}

{In this Letter, 
by using an efficient unbiased numerical method, i.e.,
the thermal pure quantum (TPQ) method~\cite{Sugiura2012},} 
we systematically study finite-temperature
properties of the frustrated Hubbard model, 
which
is a prototypical system where the 
competition between the geometrical frustrations and
the strong electronic correlations plays a crucial role.
From {the} unbiased and systematic calculations,
{we clarify how the geometrical frustrations controls the 
crossover temperatures of the Mot transitions}
and find the finite-temperature signatures of QSL.
We also propose an experimental criterion of closeness to the spin liquid phase: Finite-temperature  
entropy at moderately high temperatures 
significantly correlates with closeness to the spin liquid phase.
{Experimental searches for spin liquids have so far 
focused on setting up an {\it alibi} of spontaneous symmetry breakings
down to ultra-low temperatures.
However, we reveal that, even at moderately high temperatures $T$$\sim$$t/10$,
it becomes clear whether the target system has chance to be a spin liquid at zero temperature.}

{{\it Model and Methods.}--}
We study the $t$-$t^{\prime}$ Hubbard model on a square lattice 
(see Fig.~\ref{fig:schem})
defined as
\begin{align}
\hat{\cal H} &= 
-t\sum_{\langle i,j\rangle,\sigma}(c_{i\sigma}^{\dagger}c_{j\sigma}+\mathrm{h.c.})
-t^{\prime}\sum_{\langle\langle i,j\rangle\rangle,\sigma}(c_{i\sigma}^{\dagger}c_{j\sigma}+\mathrm{h.c.}) \notag \\ 
&+U\sum_{i}n_{i\uparrow}n_{i\downarrow},
\label{eq:H}
\end{align}
where 
$c_{i\sigma}^{\dagger}$ ($c_{i\sigma}$) is a creation (annihilation) operator of 
an electron with spin $\sigma$ at $i$th site.
The first (second) term describes the hopping of electrons between the 
{nn (nnn)} 
sites $\langle i,j \rangle$ ($\langle\langle i,j \rangle\rangle$) 
on the square lattice, 
and the third term represents the onsite Coulomb interactions ($U$$>$$0$).
In the following, we 
focus on the half filling, 
i.e., the filling is given by 
$n$$=$$N_\mathrm{s}^{-1} \sum_{i \sigma} \langle c_{i \sigma}^\dagger c_{i \sigma} \rangle$$=$$1$ 
($N_\mathrm{s}$=$L$$\times$$L$ is the system size).
To reduce the numerical cost, we only consider the total $S^{z}$$=$$0$ space, i.e.,
 $S^{z}_{\rm total}$$=$$\sum_{i}S_{i}^{z}$$=$$0$.
We employ a $4$$\times$$4$ cluster with 
{a} periodic boundary condition in the most of the present Letter~\cite{note_size}.

{In the TPQ method~\cite{Sugiura2012},
by multiplying $(l-\hat{\cal{H}}/N_{\rm s})$ to random 
vector $|\psi_{\rm rand}\rangle$,
we numerically generate the TPQ state.
Here, $l$ is constant that is larger than
the maximum eigenvalue of $\hat{\cal{H}}/N_{\rm s}$.
The $k$th TPQ state is {recursively} defined as
$|\psi_{k}\rangle\equiv{(l-\hat{\cal{H}}/N_{\rm s})|\psi_{k-1}\rangle}/{|(l-\hat{\cal{H}}/N_{\rm s})|\psi_{k-1}\rangle|}$
{with} $|\psi_{0}\rangle$=$|\psi_{\rm rand}\rangle$.
It is shown that the temperature $T_{k}$
{corresponding to}
the $k$th TPQ state is estimated from the $k$th 
internal energy {$u_{k}$$=$$\langle\psi_{k}|\hat{\cal{H}}|\psi_{k}\rangle/N_{\rm s}$} within the accuracy of $O(1/N_{\rm s})$,
{as} $\beta_{k}$$=$$1/k_{\rm B}T_{k}$$=$$2k/N_{\rm s}(l-u_{k})+O(1/N_{\rm s})$,
where $k_{\rm B}$ is the Boltzmann constant and we take $k_{\rm B}$$=$$1$ in this {Letter}.
It is shown that physical properties at $T$$=$$T_{k}$ can be calculated as the
expectation value {taken} with respect to $|\psi_{k}\rangle$, i.e.,
$\langle \hat{A} \rangle_{T=T_{k}}=\langle\psi_{k}|\hat{A}|\psi_{k}\rangle+O(1/N_{\rm s}).$
To estimate the finite-size error, we typically perform 
{five} runs {initiated with different $|\psi_{\rm rand}\rangle$} and regard its standard deviations as error bars.
{
{Here,}
note that, in the pioneering
works~\cite{Imada1986,FiniteLanczos,Hams2000},
the finite-temperature {observables were already calculated}
{by replacing ensemble average with random sampling of wave functions}.
}
}

{{\it {Finite-temperature physical quantities in Hubbard models}.}--}
{We first show the results of the finite-$T$ calculations for
$t^{\prime}/t$$=$$0.5$ as an example {of weakly frustrated Hubbard models}.
{The} ground state is expected to be N\'eel state {for $t^{\prime}/t$$=$$0.5$}.
Figure \ref{fig:tpr05} (a) shows that temperature dependence of the specific
heat $C/N_{\rm s}$, which is {given by}
${C}/{N_{\rm s}}$$=$$(\langle\hat{\cal{H}}^2\rangle-\langle\hat{\cal H}\rangle^2){/}(N_{\rm s}T^2).$
{The specific heat has {a} single peak for $U/t$$=$$4$ {as a function of $T$} while
double-peak structures~\cite{PhysRevB.5.1966}
are universal at strong-coupling regions of the Hubbard-type models
irrespective of dimensionality~\cite{PhysRevB.36.3833,KAWAKAMI1989287,PhysRevB.48.7167,PhysRevB.55.12918}.}

The high-temperature peak of {$C$}
is generated
by the charge degrees of freedom~\cite{PhysRevB.5.1966,Juttner1998471}
{whose energy scale}
is determined by $U$,
as confirmed later by the peak temperatures insensitive to $t'$ shown in Fig.~\ref{fig:SS}(c).
Below the peak temperature, the Mott gap opens and charge degrees of freedom {begin} to freeze.
In other words, below the peak temperature,
electrons begin to feel the {on-site} repulsion $U$ and
double occupancy, which is
{measured by}
$D$$=$${N_{s}^{-1}}\sum_{i=1,N_{s}}\langle n_{i\uparrow}n_{i\downarrow}\rangle$,
is gradually prohibited.

\begin{figure}[tb!]
  \begin{center}
    \includegraphics[width=9cm,clip]{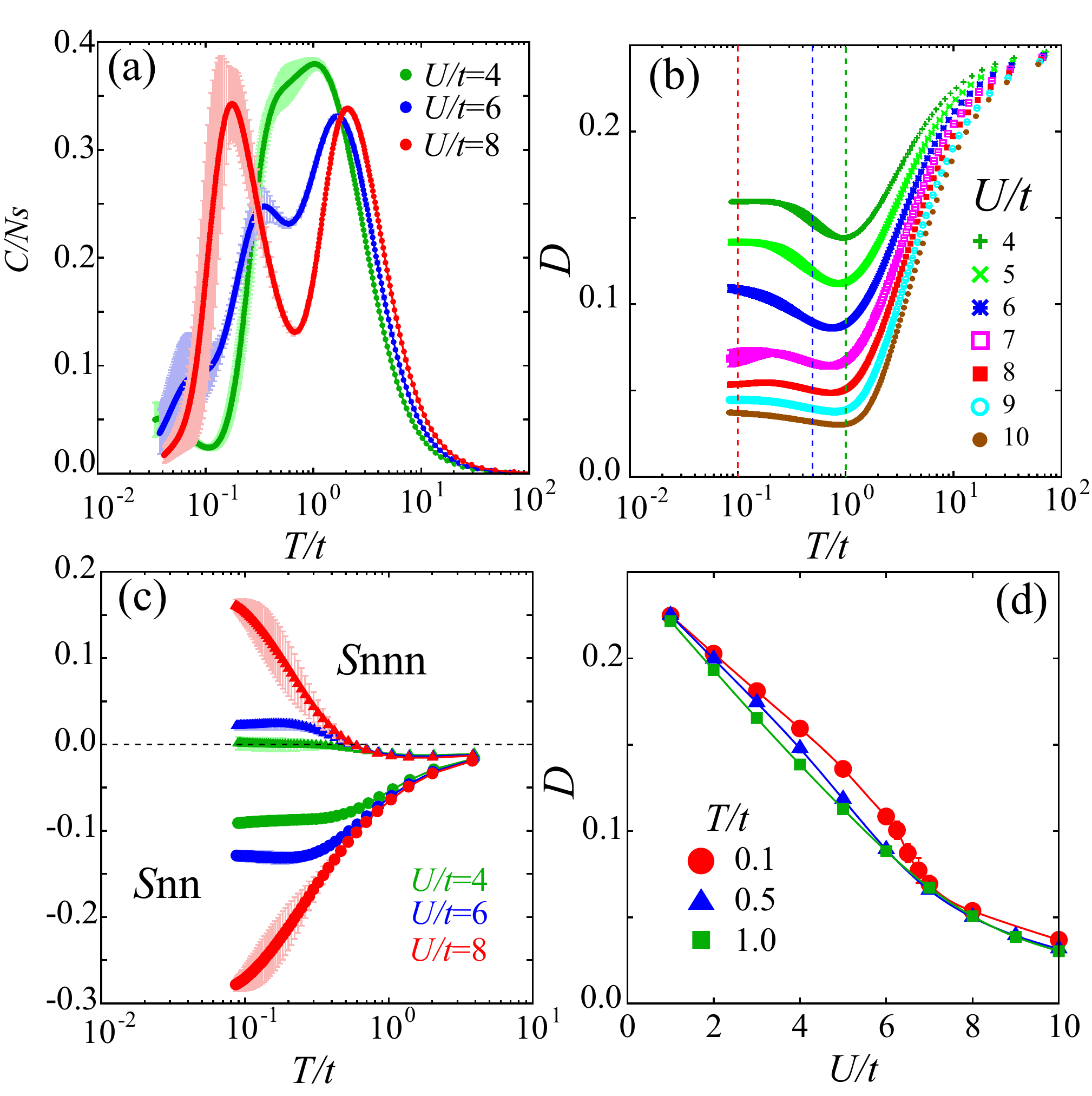}
  \end{center}
\caption{(color online).
(a)~Temperature dependence of the specific heat.
Error bars are shown as shaded regions.
(b)~Temperature dependence of double occupancy $D$ for several $U$.
Irrespective of interaction strength, we find the non-monotonic
temperature dependence of $D$.
(c)~Temperature dependence of the 
{nn (nnn) 
spin correlation $S_{\rm nn}$ ($S_{\rm nnn}$), which are defined as
{$S_{p}=1/(z_{p}N_{\rm s})\sum_{i=1}^{N_{\rm s}}\sum_{\mu}\vec{S}_{i}\cdot\vec{S}_{i+\vec{e}_\mu}$,}
where $p$=nn or nnn and accordingly,
$\vec{e}_\mu$ runs over nn or nnn sites, and
{$z_{p}$ represents the coordination number for nn or nnn sites}.}
(d)~Interaction ($U$) dependence of $D$
for $T/t=1.0,0.5,0.1$. By lowering temperature, we can see the 
signature of the finite-temperature Mott transition.
}
\label{fig:tpr05}
\end{figure}

We show temperature dependence of the double occupancy in Fig.~\ref{fig:tpr05}~(b).
Our simulation shows non-monotonic temperature dependence 
of $D$ from weak to strong coupling region, 
i.e., $D$ has the minimum around {$T/t$$\sim$$1$}. 
{We note that this non-monotonic behavior is 
universal one and observed in a wide range of Hubbard-type models~\cite{OnodaImada,Thomale,LeBlancPRX,Takai2016}.} 
The non-monotonic temperature dependence
is explained by
the development of antiferromagnetic correlations.
{Because the antiferromagnetic correlations induce
singlet states that have larger double occupancies 
compared to the other states,
the double occupancy
increases at low temperature.}

The low-temperature peak of the specific heat, in contrast to the high-temperature peak, 
is induced by
spin degrees of freedom{~\cite{PhysRevB.5.1966,PhysRevB.36.3833,KAWAKAMI1989287,PhysRevB.48.7167,PhysRevB.55.12918}.}
{This}
signals development of
antiferromagnetic correlations as shown in Fig.~\ref{fig:tpr05}~(c),
which corresponds to {an} increase in $D$ at low temperatures as discussed above.
The peak temperature becomes lower 
as $U$ increases,
as is manifest in Fig.~\ref{fig:tpr05}~(a), which
is consistent with
the characteristic energy scale of the spin-degrees of 
freedom at the strong coupling limit
given by the effective superexchange $J_{1}\sim 4t^2/U$.
To examine the energy scale, we show
the temperature dependence of the
{nn and  nnn} spin correlations in Fig.~\ref{fig:tpr05}~(c).
As it is expected,
the spin correlations develop around the
low-temperature peak of the specific heat.
The two emergent energy scales
corresponding to spin and charge degrees of freedom
in the strong coupling region
are indeed identified as 
origin of two-peak structure of the specific heat for $U/t$$\gtrsim$$6$
{while separation of these energy scales may not be clear in excitation spectra~\cite{PhysRevLett.105.267204}}.

{\it Effect of geometrical frustration on metal-insulator transitions.}--
To examine the signature of
the finite-temperature Mott transitions, we 
calculate the $U$-dependence of $D$ for several temperatures as shown in
Fig.~\ref{fig:tpr05}~(d).
By lowering the temperatures,
we find the slope of $D$ becomes steep.
{Since} the slope of $D$ diverges at the finite-temperature
Mott critical end point~\cite{Kotliar2000},
this behavior 
{can be regarded as the crossover of the finite-temperature 
Mott critical point.}

\begin{figure}[b!]
  \begin{center}
    \includegraphics[width=8.5cm,clip]{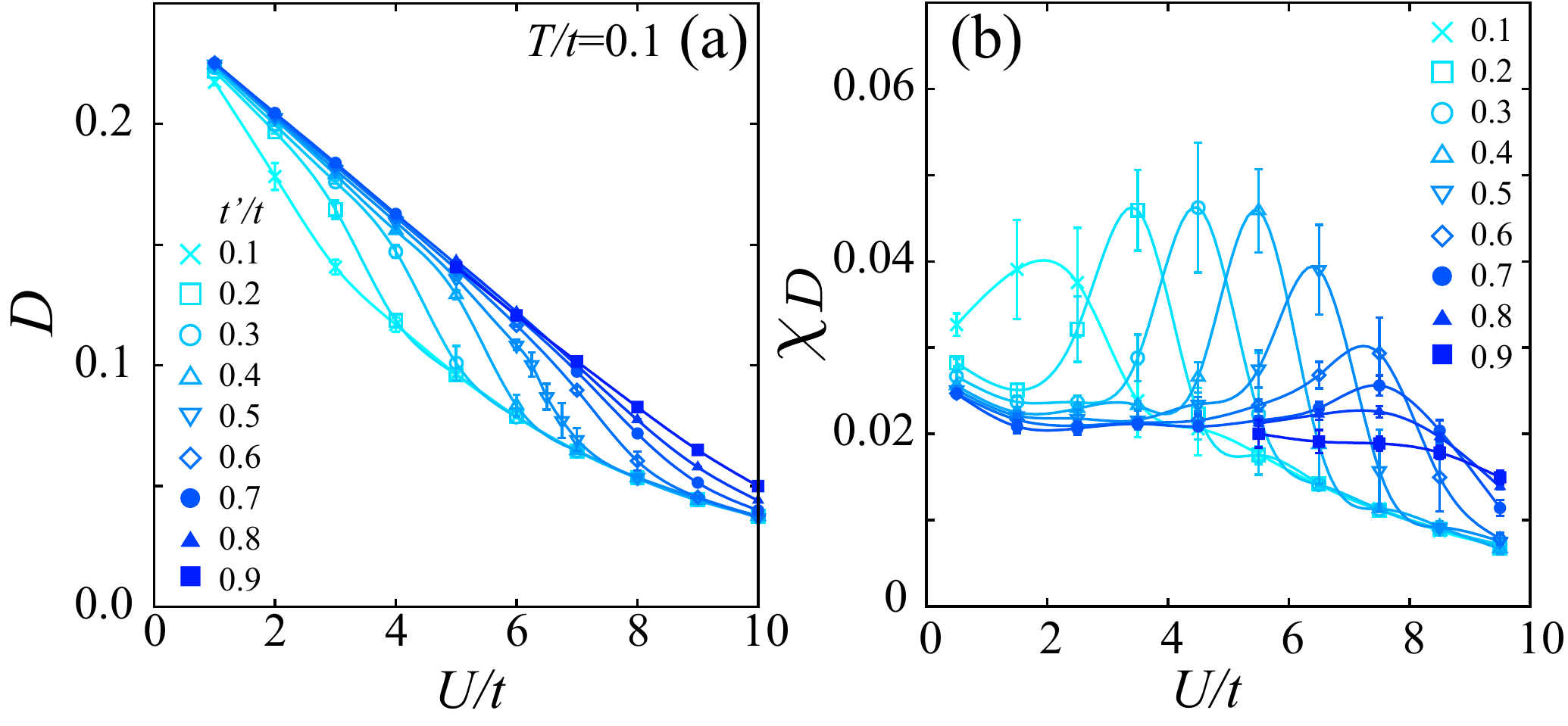}
  \end{center}
\caption{(color online).
(a)~Interaction dependence of double occupancy $D$ for several 
$t'$ at $T/t=0.1$.
The crossover interaction becomes larger by increasing frustration.
(b)~Interaction dependence of double occupancy susceptibility $\chi_{D}$
for several $t'$.
{To reduce the numerical error in numerical differentiation,
we take finite difference $\Delta U=1$.}
We see the height of peak in $\chi_{D}$ becomes smaller by
increasing $t^{\prime}$.
This result indicates that the critical temperature of
Mott critical end point becomes lower by increasing $t^{\prime}$.
{Solid curves are guides for eyes.}
}
\label{fig:chiD}
\end{figure}

{
To see the $t^{\prime}$ dependence of
{the critical temperatures of}
the Mott transitions 
{from the crossover behaviors at fixed temperature},
we calculate the $U$ dependence of $D$ for several different 
$t^{\prime}$ at $T/t=0.1$ as shown in Fig.~\ref{fig:chiD}~(a).
From this data, by performing the numerical differentiation for $D$
with respect to $U$, 
we obtain doublon susceptibility $\chi_{D}=-\partial D/\partial U$.
{Here, note that, even
at zero temperature, the maxima of $\chi_{D}$ as the function of $U/t$
have been demonstrated to signal the Mott transitions in 
the finite-size Hubbard models~\cite{doi:10.1143/JPSJ.76.074719}.}
The obtained $\chi_{D}$ is shown in Fig.~\ref{fig:chiD}(b).
{By increasing $t^{\prime}$ (increasing frustration),
we find that the peak values of $\chi_{D}$ {at fixed temperature} decrease and 
the peak almost vanishes around $t^{\prime}/t\sim0.75$.}
This result indicates that
{the critical temperature of the critical end point of the Mott transitions}
becomes lower by increasing the frustration and
the marginal quantum critical point (MQCP)~\cite{ImadaMQCP,MisawaImada} 
{exists} around $t^{\prime}/t\sim 0.75$,
where the critical temperature 
of the Mott transition becomes zero.
{Because of} the limitation of the available system size,
it is hard to
{make a conclusion to the} fate of the finite-temperature 
Mott critical point.
However, our results are qualitatively consistent with
the mean-field calculations~\cite{misawa2006,MisawaImada} and
it is plausible that the MQCP
appears around $t^{\prime}/t\sim 0.75$.
{We note that ground-state calculations for the Hubbard model on 
the  anisotropic triangular lattice
also indicate that nature of Mott transitions 
is governed by the geometrical 
frustrations~\cite{MoritaPIRG,ImadaMQCP}.}

{{\it Signatures of QSL}.--}
Here, to examine the signature of the 
spin liquid state, we calculate spin correlations
for several different $t^{\prime}$.
As shown in Fig.~\ref{fig:SS}(a) and (b),
in the small $t^{\prime}$ ($t^{\prime}\lesssim 0.6$),
by lowering the temperature,
antiferromagnetic {{\rm nn}} 
spin correlations develop
while the ferromagnetic {{\rm nnn}} 
spin correlations develop.
These spin correlations are consistent with the N\'eel order.
In contrast to this, for large $t^{\prime}$ region ($t^{\prime}/t\gtrsim 0.8$),
{while} the antiferromagnetic {{\rm nnn}} 
spin correlations develop,
{the {\rm nn}} 
spin correlations {remain} small {even at low temperatures below $t/10$}.
These spin correlations indicate that the stripe antiferromagnetic order becomes
stable in the {large} $t^{\prime}$ region.
Sandwiched by the N\'eel and the stripe {orders},
$S_{\rm nnn}$
{is saturated and remains small even at low temperatures for the intermediate $t^{\prime}$.
{We note that short-range spin correlations 
at moderately high temperatures ($T/t\sim 0.1$) 
reflect the corresponding ground states and the
behavior at $t^{\prime}/t=0.75$ is consistent with that of the QSL.}
}

We next examine the thermodynamic properties 
of the 
{spin-liquid candidates}.
In contrast to {the high-temperature peaks of $C$ insensitive to $t^{\prime}$ shown in Fig.~\ref{fig:SS}(c)},
the positions of the second peak largely depend 
on 
$t^{\prime}$
{since} they are governed by
the spin degrees of freedom.  
At the {highly frustrated parameter region $t^{\prime}/t$\tc{$\sim$}$0.75$,} 
the amplitude of the second peak {remains} small and
indicates the {substantial amount of low-energy} excitations 
{is left even below the energy scale} $T/t$$\sim$$0.05$.

\begin{figure}[t!]
  \begin{center}
    \includegraphics[width=9.5cm,clip]{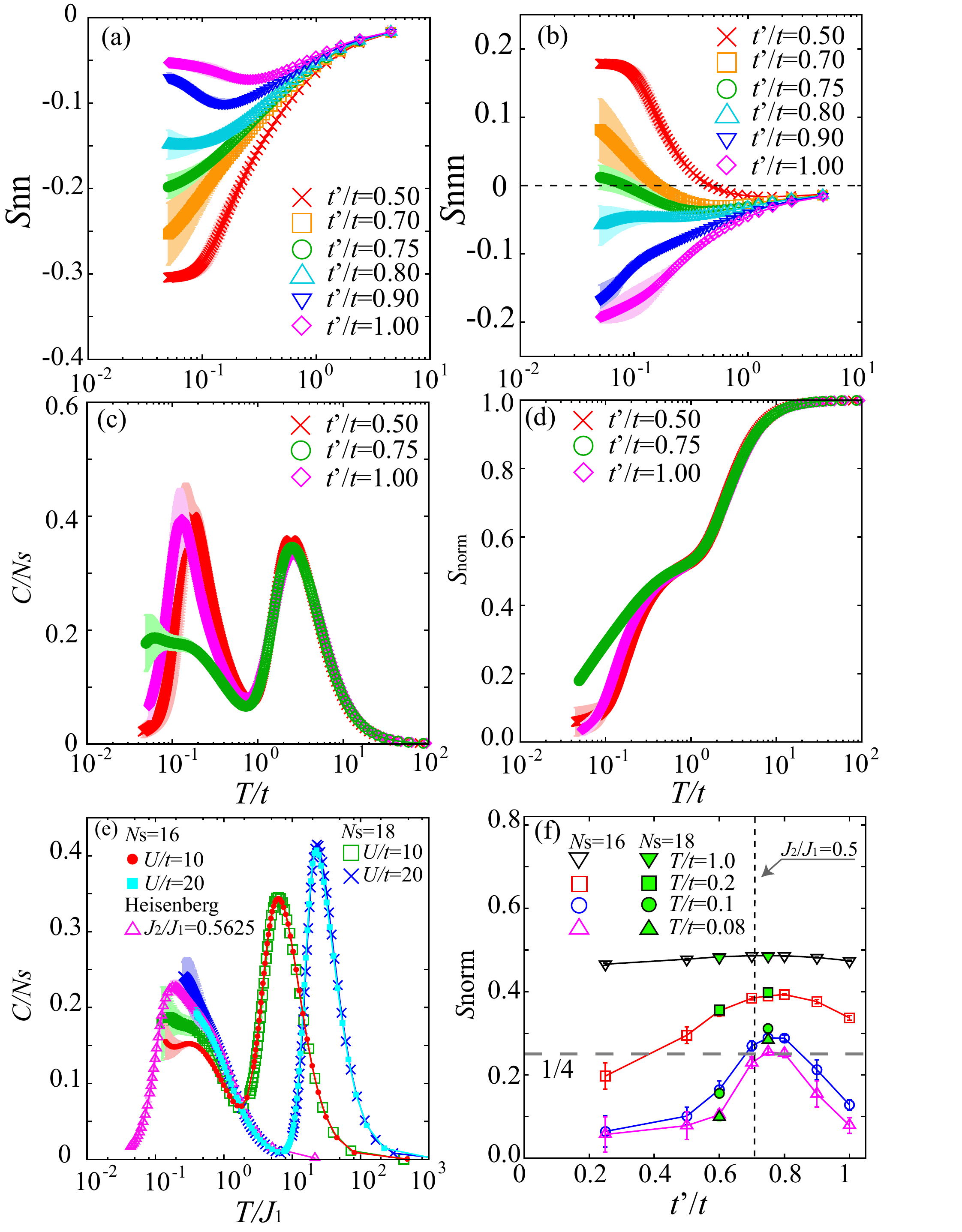}
  \end{center}
\caption{(color online).
(a),(b)~
Temperature dependence of nearest-neighbor ($S_{\rm nn}$) and next-nearest-neighbor ($S_{\rm nnn}$)
spin correlations for several different $t^{\prime}$ at $U/t=10$.
Around $t^{\prime}\sim0.75$, $S_{\rm nnn}$ becomes almost zero at low temperature.
(c)~Temperature dependence of the specific heat for several $t^{\prime}$ at $U/t=10$.
(d)~Temperature dependence of the normalized entropy, 
{which is defined by
$S(T)/N_{\rm s}=c\ln{2}-1/N_{\rm s}\int^{T}_{\infty}C/TdT$,~
$S_{\rm norm}$$=$$S/(c\ln{2})$, 
where constant $c$ is 2 if 
{$S^{z}_{\rm total}$ is unrestricted}. 
In this calculation, we 
{set} {$S^{z}_{\rm total}$$=$$0$.} 
{Then,} $c$ is given by $c= \ln{(_{16}C_{8})^2}/16\ln{2}\sim 1.706$ for 16 sites.
If we take {the thermodynamic limit}, $c$ {converges} to 2.}
For $t^{\prime}/t\sim0.75$, large residual entropy is observed compared to $t^{\prime}/t=0.5, 1.0$. 
\tc{(e)~Temperature dependence of the specific heat of $4\times 4$ and $3\sqrt{2}\times 3\sqrt{2}$ Hubbard clusters
for $t'=0.75$ at $U/t=10$ and $20$ compared with a $4\sqrt{2}\times 4\sqrt{2}$ $J_1$-$J_2$ Heisenberg cluster~\cite{note_size}
by setting $J_1=4t^2/U$ and $J_2=4{t'}^2/U$.}
(f)~\tc{Frustration ($t^{\prime}/t$)} dependence of the entropy for several \tc{fixed} temperatures.
Around $t^{\prime}/t\sim0.75$,
the large remaining entropy is observed
and it is the evidence of QSL.
}
\label{fig:SS}
\end{figure}

{To 
{quantify the amount of the low-energy excitations}, 
we calculate the entropy. 
We show temperature dependence of $S_{\rm norm}$
in Fig~\ref{fig:SS}(d).
At the {highly frustrated} region ($t^{\prime}/t$$=$$0.75$),
the entropy is not released down to $T/t$$\sim$$0.05$ compared to
weakly frustrated {regions}.

{To examine the finite-size effects,
{we show the specific heat of the 18-site cluster compared 
with that for the larger $U/t$ and the strong
 coupling limit~\cite{note_size} {in Fig.~4(e)}
by setting $J_1 = 4t^2/U$ and $J_2 = 4{t'}^2/U$, at the highly frustrated parameter $t'/t=0.75$.}
Although small system size dependence exists,
{all the data consistently shows that the 
reduction of the low-temperature-peak height occurs at $t^{\prime}/t=0.75$
compared to the region where the magnetic long-range orders appear.}
}

We also show $t^{\prime}$ dependence of the entropy
\tc{at} several \tc{fixed} temperature\tc{s}
in Fig.~\ref{fig:SS}(f).
{We find that the entropy \tc{at fixed temperature} has peak
around
\tc{$t'/t\sim 0.75$, where}
the spin-liquid or non-magnetic ground states
\tc{are expected} at the strong coupling limit~\cite{Gong2014,Morita2015}.}
In sharp contrast, for the N\'eel and stripe order, {the entropy} 
{quickly} becomes zero by decreasing the temperature, which
indicates that {almost} all the degrees of {freedom} 
including 
spin degrees of freedom is released below $T$$\sim$$t/10$.

}

{Even at the moderately high temperatures $T$$\sim$$t/10$, therefore,
{the entropy} 
clearly shows whether the target systems have 
chance to be spin-liquid states at zero temperature.
This fact is seemingly trivial since, in the presence of the geometrical frustrations,} entropy is expected to 
remain finite at low temperatures well below
the exchange coupling $J_1$. 
However, the present result offers the first unbiased and quantitative 
criterion for the emergence of the spin-liquid ground states 
{in the geometrically frustrated Mott insulators}.
{Although competition among quantum phases is also expected to show remaining entropy,
there are counterexamples.
An example is
the quantum phase transition from the Kitaev spin liquid~\cite{AnnalsofPhysics321.2} to ordered states~\cite{trebst2017kitaev}.
When the ground state changes from the spin liquid to an ordered state,
entropy at a temperature equal to, for example, one-quarter of the dominant energy scale
monotonically decreases and does not show any enhancement above the
transition point~\cite{PhysRevB.93.174425}.}

In summary,
we apply the TPQ method to the frustrated Hubbard model.
By calculating the susceptibilities of the double occupancy,
we find that the characteristic energy scale of the Mott transition 
becomes lower by increasing $t^{\prime}$.
This result indicates emergence of the MQCP around $t^{\prime}/t\sim0.75$.
{We note that the MQCP and QSL appear around nearly the 
same parameter region and we expect that this coincide is not accidental one:
Since infinitesimally small antiferromagnetic 
order parameters cannot generate a single-particle gap on the 
entire Fermi surface, another exotic phases such as the QSL are expected to appear~\cite{Senthil,MPAFisher}
in between the paramagnetic metal and the antiferromagnetic insulator 
if the metal-insulator transition is continuous. We note that it is 
unlikely but not excluded that the antiferromagnetic 
metal appears as the intermediate phase as in the 
mean-field calculations~\cite{misawa2006,MisawaImada}. It is an intriguing issue 
left for future studies to examine whether the QSL 
universally appears around the continuous metal insulator transitions including MQCP or not.
{We also find \tc{that large entropy remains at $T/t\sim 0.1$}
around the spin-liquid \tc{region} and
this may be useful criterion for searching the QSL.
We note that the suppression of magnetic orders due to the 
low dimensionality~\cite{PhysRevB.68.094423} and impurities~\cite{doi:10.7566/JPSJ.83.034714},
instead of the geometrical frustration, do not induces the large remaining entropy. 
For the spin liquid candidates, 
it is an intriguing challenge 
to examine whether the proposal will work.}

\begin{acknowledgements}
The authors thank Masatoshi Imada for fruitful disccusions and comments.
A part of calculations is done by using open-source software H$\Phi$~\cite{hphi,hphi_ma,hphi_git}.
Our calculation was partly carried out at the 
Supercomputer Center, Institute for Solid State Physics, University of Tokyo.
This work was supported by JSPS KAKENHI ({Grant Nos.~15K17702,16K17746, and 16H06345})
{and was supported by PRESTO, JST}.
{This work was also supported in part by MEXT as a social and 
scientific priority issue (Creation of new functional devices and high-performance materials
to support next-generation industries) to be tackled by using post-K computer.}
TM was supported by Building of Consortia for the Development
of Human Resources in Science and Technology from the MEXT of Japan.
\end{acknowledgements}


\begin{thebibliography}{49}
\expandafter\ifx\csname natexlab\endcsname\relax\def\natexlab#1{#1}\fi
\expandafter\ifx\csname bibnamefont\endcsname\relax
  \def\bibnamefont#1{#1}\fi
\expandafter\ifx\csname bibfnamefont\endcsname\relax
  \def\bibfnamefont#1{#1}\fi
\expandafter\ifx\csname citenamefont\endcsname\relax
  \def\citenamefont#1{#1}\fi
\expandafter\ifx\csname url\endcsname\relax
  \def\url#1{\texttt{#1}}\fi
\expandafter\ifx\csname urlprefix\endcsname\relax\def\urlprefix{URL }\fi
\providecommand{\bibinfo}[2]{#2}
\providecommand{\eprint}[2][]{\url{#2}}

\bibitem[{\citenamefont{Imada et~al.}(1998)\citenamefont{Imada, Fujimori, and
  Tokura}}]{ImadaRMP}
\bibinfo{author}{\bibfnamefont{M.}~\bibnamefont{Imada}},
  \bibinfo{author}{\bibfnamefont{A.}~\bibnamefont{Fujimori}}, \bibnamefont{and}
  \bibinfo{author}{\bibfnamefont{Y.}~\bibnamefont{Tokura}},
  \bibinfo{journal}{Rev. Mod. Phys.} \textbf{\bibinfo{volume}{70}},
  \bibinfo{pages}{1039} (\bibinfo{year}{1998}).

\bibitem[{\citenamefont{Kanoda}(2006)}]{Kanoda}
\bibinfo{author}{\bibfnamefont{K.}~\bibnamefont{Kanoda}}, \bibinfo{journal}{J.
  Phys. Soc. Jpn.} \textbf{\bibinfo{volume}{75}}, \bibinfo{pages}{051007}
  (\bibinfo{year}{2006}).

\bibitem[{\citenamefont{Bloch et~al.}(2008)\citenamefont{Bloch, Dalibard, and
  Zwerger}}]{BlochRMP}
\bibinfo{author}{\bibfnamefont{I.}~\bibnamefont{Bloch}},
  \bibinfo{author}{\bibfnamefont{J.}~\bibnamefont{Dalibard}}, \bibnamefont{and}
  \bibinfo{author}{\bibfnamefont{W.}~\bibnamefont{Zwerger}},
  \bibinfo{journal}{Rev. Mod. Phys.} \textbf{\bibinfo{volume}{80}},
  \bibinfo{pages}{885} (\bibinfo{year}{2008}).

\bibitem[{Die()}]{Diep}
\bibinfo{note}{For instance, Frustrated Spin Systems, ed. H. Diep (World
  Scientific, Singapore, 2005).}

\bibitem[{\citenamefont{Shannon et~al.}(2012)\citenamefont{Shannon, Sikora,
  Pollmann, Penc, and Fulde}}]{Shannon_PRL2012}
\bibinfo{author}{\bibfnamefont{N.}~\bibnamefont{Shannon}},
  \bibinfo{author}{\bibfnamefont{O.}~\bibnamefont{Sikora}},
  \bibinfo{author}{\bibfnamefont{F.}~\bibnamefont{Pollmann}},
  \bibinfo{author}{\bibfnamefont{K.}~\bibnamefont{Penc}}, \bibnamefont{and}
  \bibinfo{author}{\bibfnamefont{P.}~\bibnamefont{Fulde}},
  \bibinfo{journal}{Phys. Rev. Lett.} \textbf{\bibinfo{volume}{108}},
  \bibinfo{pages}{067204} (\bibinfo{year}{2012}).

\bibitem[{\citenamefont{Normand and Nussinov}(2014)}]{Normand_PRL2014}
\bibinfo{author}{\bibfnamefont{B.}~\bibnamefont{Normand}} \bibnamefont{and}
  \bibinfo{author}{\bibfnamefont{Z.}~\bibnamefont{Nussinov}},
  \bibinfo{journal}{Phys. Rev. Lett.} \textbf{\bibinfo{volume}{112}},
  \bibinfo{pages}{207202} (\bibinfo{year}{2014}).

\bibitem[{\citenamefont{Shimizu et~al.}(2003)\citenamefont{Shimizu, Miyagawa,
  Kanoda, Maesato, and Saito}}]{Shimizu2003}
\bibinfo{author}{\bibfnamefont{Y.}~\bibnamefont{Shimizu}},
  \bibinfo{author}{\bibfnamefont{K.}~\bibnamefont{Miyagawa}},
  \bibinfo{author}{\bibfnamefont{K.}~\bibnamefont{Kanoda}},
  \bibinfo{author}{\bibfnamefont{M.}~\bibnamefont{Maesato}}, \bibnamefont{and}
  \bibinfo{author}{\bibfnamefont{G.}~\bibnamefont{Saito}},
  \bibinfo{journal}{Phys. Rev. Lett.} \textbf{\bibinfo{volume}{91}},
  \bibinfo{pages}{107001} (\bibinfo{year}{2003}).

\bibitem[{\citenamefont{Itou et~al.}(2008)\citenamefont{Itou, Oyamada, Maegawa,
  Tamura, and Kato}}]{Itou2008}
\bibinfo{author}{\bibfnamefont{T.}~\bibnamefont{Itou}},
  \bibinfo{author}{\bibfnamefont{A.}~\bibnamefont{Oyamada}},
  \bibinfo{author}{\bibfnamefont{S.}~\bibnamefont{Maegawa}},
  \bibinfo{author}{\bibfnamefont{M.}~\bibnamefont{Tamura}}, \bibnamefont{and}
  \bibinfo{author}{\bibfnamefont{R.}~\bibnamefont{Kato}},
  \bibinfo{journal}{Phys. Rev. B} \textbf{\bibinfo{volume}{77}},
  \bibinfo{pages}{104413} (\bibinfo{year}{2008}).

\bibitem[{\citenamefont{Kanoda and Kato}(2011)}]{kanoda2011mott}
\bibinfo{author}{\bibfnamefont{K.}~\bibnamefont{Kanoda}} \bibnamefont{and}
  \bibinfo{author}{\bibfnamefont{R.}~\bibnamefont{Kato}},
  \bibinfo{journal}{Annu. Rev. Condens. Matter Phys.}
  \textbf{\bibinfo{volume}{2}}, \bibinfo{pages}{167} (\bibinfo{year}{2011}).

\bibitem[{\citenamefont{Balents}(2010)}]{balents2010spin}
\bibinfo{author}{\bibfnamefont{L.}~\bibnamefont{Balents}},
  \bibinfo{journal}{Nature} \textbf{\bibinfo{volume}{464}},
  \bibinfo{pages}{199} (\bibinfo{year}{2010}).

\bibitem[{\citenamefont{Kashima and Imada}(2001)}]{kashima2001magnetic}
\bibinfo{author}{\bibfnamefont{T.}~\bibnamefont{Kashima}} \bibnamefont{and}
  \bibinfo{author}{\bibfnamefont{M.}~\bibnamefont{Imada}}, \bibinfo{journal}{J.
  Phys. Soc. Jpn.} \textbf{\bibinfo{volume}{70}}, \bibinfo{pages}{3052}
  (\bibinfo{year}{2001}).

\bibitem[{\citenamefont{Mizusaki and Imada}(2006)}]{MizusakiImada}
\bibinfo{author}{\bibfnamefont{T.}~\bibnamefont{Mizusaki}} \bibnamefont{and}
  \bibinfo{author}{\bibfnamefont{M.}~\bibnamefont{Imada}},
  \bibinfo{journal}{Phys. Rev. B} \textbf{\bibinfo{volume}{74}},
  \bibinfo{pages}{014421} (\bibinfo{year}{2006}).

\bibitem[{\citenamefont{Tocchio et~al.}(2008)\citenamefont{Tocchio, Becca,
  Parola, and Sorella}}]{Tocchio2008}
\bibinfo{author}{\bibfnamefont{L.~F.} \bibnamefont{Tocchio}},
  \bibinfo{author}{\bibfnamefont{F.}~\bibnamefont{Becca}},
  \bibinfo{author}{\bibfnamefont{A.}~\bibnamefont{Parola}}, \bibnamefont{and}
  \bibinfo{author}{\bibfnamefont{S.}~\bibnamefont{Sorella}},
  \bibinfo{journal}{Phys. Rev. B} \textbf{\bibinfo{volume}{78}},
  \bibinfo{pages}{041101} (\bibinfo{year}{2008}).

\bibitem[{\citenamefont{Jiang et~al.}(2012)\citenamefont{Jiang, Yao, and
  Balents}}]{Jiang2012}
\bibinfo{author}{\bibfnamefont{H.-C.} \bibnamefont{Jiang}},
  \bibinfo{author}{\bibfnamefont{H.}~\bibnamefont{Yao}}, \bibnamefont{and}
  \bibinfo{author}{\bibfnamefont{L.}~\bibnamefont{Balents}},
  \bibinfo{journal}{Phys. Rev. B} \textbf{\bibinfo{volume}{86}},
  \bibinfo{pages}{024424} (\bibinfo{year}{2012}).

\bibitem[{\citenamefont{Hu et~al.}(2013)\citenamefont{Hu, Becca, Parola, and
  Sorella}}]{Hu2013}
\bibinfo{author}{\bibfnamefont{W.-J.} \bibnamefont{Hu}},
  \bibinfo{author}{\bibfnamefont{F.}~\bibnamefont{Becca}},
  \bibinfo{author}{\bibfnamefont{A.}~\bibnamefont{Parola}}, \bibnamefont{and}
  \bibinfo{author}{\bibfnamefont{S.}~\bibnamefont{Sorella}},
  \bibinfo{journal}{Phys. Rev. B} \textbf{\bibinfo{volume}{88}},
  \bibinfo{pages}{060402} (\bibinfo{year}{2013}).

\bibitem[{\citenamefont{Gong et~al.}(2014)\citenamefont{Gong, Zhu, Sheng,
  Motrunich, and Fisher}}]{Gong2014}
\bibinfo{author}{\bibfnamefont{S.-S.} \bibnamefont{Gong}},
  \bibinfo{author}{\bibfnamefont{W.}~\bibnamefont{Zhu}},
  \bibinfo{author}{\bibfnamefont{D.~N.} \bibnamefont{Sheng}},
  \bibinfo{author}{\bibfnamefont{O.~I.} \bibnamefont{Motrunich}},
  \bibnamefont{and} \bibinfo{author}{\bibfnamefont{M.~P.~A.}
  \bibnamefont{Fisher}}, \bibinfo{journal}{Phys. Rev. Lett.}
  \textbf{\bibinfo{volume}{113}}, \bibinfo{pages}{027201}
  (\bibinfo{year}{2014}).

\bibitem[{\citenamefont{Morita et~al.}(2015)\citenamefont{Morita, Kaneko, and
  Imada}}]{Morita2015}
\bibinfo{author}{\bibfnamefont{S.}~\bibnamefont{Morita}},
  \bibinfo{author}{\bibfnamefont{R.}~\bibnamefont{Kaneko}}, \bibnamefont{and}
  \bibinfo{author}{\bibfnamefont{M.}~\bibnamefont{Imada}}, \bibinfo{journal}{J.
  Phys. Soc. Jpn.} \textbf{\bibinfo{volume}{84}}, \bibinfo{pages}{024720}
  (\bibinfo{year}{2015}).

\bibitem[{\citenamefont{Sugiura and Shimizu}(2012)}]{Sugiura2012}
\bibinfo{author}{\bibfnamefont{S.}~\bibnamefont{Sugiura}} \bibnamefont{and}
  \bibinfo{author}{\bibfnamefont{A.}~\bibnamefont{Shimizu}},
  \bibinfo{journal}{Phys. Rev. Lett.} \textbf{\bibinfo{volume}{108}},
  \bibinfo{pages}{240401} (\bibinfo{year}{2012}).

\bibitem[{not()}]{note_size}
\bibinfo{note}{At the highly frustrated region ($t'/t\sim 0.75$), we also show
  results for a 18-site ($3\sqrt{2}\times 3\sqrt{2}$) cluster with the
  anti-periodic boundary condition and a 32-site ($4\sqrt{2}\times 4\sqrt{2}$)
  cluster of the strong coupling limit with the periodic boundary condition to
  examine the finite-size effects.}

\bibitem[{\citenamefont{Imada and Takahashi}(1986)}]{Imada1986}
\bibinfo{author}{\bibfnamefont{M.}~\bibnamefont{Imada}} \bibnamefont{and}
  \bibinfo{author}{\bibfnamefont{M.}~\bibnamefont{Takahashi}},
  \bibinfo{journal}{J. Phys. Soc. Jpn.} \textbf{\bibinfo{volume}{55}},
  \bibinfo{pages}{3354} (\bibinfo{year}{1986}).

\bibitem[{\citenamefont{Jakli\ifmmode~\check{c}\else \v{c}\fi{} and
  Prelov\ifmmode~\check{s}\else \v{s}\fi{}ek}(1994)}]{FiniteLanczos}
\bibinfo{author}{\bibfnamefont{J.}~\bibnamefont{Jakli\ifmmode~\check{c}\else
  \v{c}\fi{}}} \bibnamefont{and}
  \bibinfo{author}{\bibfnamefont{P.}~\bibnamefont{Prelov\ifmmode~\check{s}\else
  \v{s}\fi{}ek}}, \bibinfo{journal}{Phys. Rev. B}
  \textbf{\bibinfo{volume}{49}}, \bibinfo{pages}{5065} (\bibinfo{year}{1994}).

\bibitem[{\citenamefont{Hams and De~Raedt}(2000)}]{Hams2000}
\bibinfo{author}{\bibfnamefont{A.}~\bibnamefont{Hams}} \bibnamefont{and}
  \bibinfo{author}{\bibfnamefont{H.}~\bibnamefont{De~Raedt}},
  \bibinfo{journal}{Phys. Rev. E} \textbf{\bibinfo{volume}{62}},
  \bibinfo{pages}{4365} (\bibinfo{year}{2000}).

\bibitem[{\citenamefont{Shiba and Pincus}(1972)}]{PhysRevB.5.1966}
\bibinfo{author}{\bibfnamefont{H.}~\bibnamefont{Shiba}} \bibnamefont{and}
  \bibinfo{author}{\bibfnamefont{P.~A.} \bibnamefont{Pincus}},
  \bibinfo{journal}{Phys. Rev. B} \textbf{\bibinfo{volume}{5}},
  \bibinfo{pages}{1966} (\bibinfo{year}{1972}).

\bibitem[{\citenamefont{Fye and Scalettar}(1987)}]{PhysRevB.36.3833}
\bibinfo{author}{\bibfnamefont{R.~M.} \bibnamefont{Fye}} \bibnamefont{and}
  \bibinfo{author}{\bibfnamefont{R.~T.} \bibnamefont{Scalettar}},
  \bibinfo{journal}{Phys. Rev. B} \textbf{\bibinfo{volume}{36}},
  \bibinfo{pages}{3833} (\bibinfo{year}{1987}).

\bibitem[{\citenamefont{Kawakami et~al.}(1989)\citenamefont{Kawakami, Usuki,
  and Okiji}}]{KAWAKAMI1989287}
\bibinfo{author}{\bibfnamefont{N.}~\bibnamefont{Kawakami}},
  \bibinfo{author}{\bibfnamefont{T.}~\bibnamefont{Usuki}}, \bibnamefont{and}
  \bibinfo{author}{\bibfnamefont{A.}~\bibnamefont{Okiji}},
  \bibinfo{journal}{Physics Letters A} \textbf{\bibinfo{volume}{137}},
  \bibinfo{pages}{287 } (\bibinfo{year}{1989}).

\bibitem[{\citenamefont{Georges and Krauth}(1993)}]{PhysRevB.48.7167}
\bibinfo{author}{\bibfnamefont{A.}~\bibnamefont{Georges}} \bibnamefont{and}
  \bibinfo{author}{\bibfnamefont{W.}~\bibnamefont{Krauth}},
  \bibinfo{journal}{Phys. Rev. B} \textbf{\bibinfo{volume}{48}},
  \bibinfo{pages}{7167} (\bibinfo{year}{1993}).

\bibitem[{\citenamefont{Duffy and Moreo}(1997)}]{PhysRevB.55.12918}
\bibinfo{author}{\bibfnamefont{D.}~\bibnamefont{Duffy}} \bibnamefont{and}
  \bibinfo{author}{\bibfnamefont{A.}~\bibnamefont{Moreo}},
  \bibinfo{journal}{Phys. Rev. B} \textbf{\bibinfo{volume}{55}},
  \bibinfo{pages}{12918} (\bibinfo{year}{1997}).

\bibitem[{\citenamefont{Juttner
  et~al.}(1998)\citenamefont{Juttner, Klmper, and
  Suzuki}}]{Juttner1998471}
\bibinfo{author}{\bibfnamefont{G.}~\bibnamefont{J${\rm \ddot{u}}$ttner}},
  \bibinfo{author}{\bibfnamefont{A.}~\bibnamefont{Klmper}}, \bibnamefont{and}
  \bibinfo{author}{\bibfnamefont{J.}~\bibnamefont{Suzuki}},
  \bibinfo{journal}{Nuclear Physics B} \textbf{\bibinfo{volume}{522}},
  \bibinfo{pages}{471 } (\bibinfo{year}{1998}).

\bibitem[{\citenamefont{Onoda and Imada}(2003)}]{OnodaImada}
\bibinfo{author}{\bibfnamefont{S.}~\bibnamefont{Onoda}} \bibnamefont{and}
  \bibinfo{author}{\bibfnamefont{M.}~\bibnamefont{Imada}},
  \bibinfo{journal}{Phys. Rev. B} \textbf{\bibinfo{volume}{67}},
  \bibinfo{pages}{161102} (\bibinfo{year}{2003}).

\bibitem[{\citenamefont{Laubach et~al.}(2015)\citenamefont{Laubach, Thomale,
  Platt, Hanke, and Li}}]{Thomale}
\bibinfo{author}{\bibfnamefont{M.}~\bibnamefont{Laubach}},
  \bibinfo{author}{\bibfnamefont{R.}~\bibnamefont{Thomale}},
  \bibinfo{author}{\bibfnamefont{C.}~\bibnamefont{Platt}},
  \bibinfo{author}{\bibfnamefont{W.}~\bibnamefont{Hanke}}, \bibnamefont{and}
  \bibinfo{author}{\bibfnamefont{G.}~\bibnamefont{Li}}, \bibinfo{journal}{Phys.
  Rev. B} \textbf{\bibinfo{volume}{91}}, \bibinfo{pages}{245125}
  (\bibinfo{year}{2015}).

\bibitem[{\citenamefont{LeBlanc et~al.}(2015)\citenamefont{LeBlanc, Antipov,
  Becca, Bulik, Chan, Chung, Deng, Ferrero, Henderson, Jim\'enez-Hoyos
  et~al.}}]{LeBlancPRX}
\bibinfo{author}{\bibfnamefont{J.~P.~F.} \bibnamefont{LeBlanc}},
  \bibinfo{author}{\bibfnamefont{A.~E.} \bibnamefont{Antipov}},
  \bibinfo{author}{\bibfnamefont{F.}~\bibnamefont{Becca}},
  \bibinfo{author}{\bibfnamefont{I.~W.} \bibnamefont{Bulik}},
  \bibinfo{author}{\bibfnamefont{G.~K.-L.} \bibnamefont{Chan}},
  \bibinfo{author}{\bibfnamefont{C.-M.} \bibnamefont{Chung}},
  \bibinfo{author}{\bibfnamefont{Y.}~\bibnamefont{Deng}},
  \bibinfo{author}{\bibfnamefont{M.}~\bibnamefont{Ferrero}},
  \bibinfo{author}{\bibfnamefont{T.~M.} \bibnamefont{Henderson}},
  \bibinfo{author}{\bibfnamefont{C.~A.} \bibnamefont{Jim\'enez-Hoyos}},
  \bibnamefont{et~al.} (\bibinfo{collaboration}{Simons Collaboration on the
  Many-Electron Problem}), \bibinfo{journal}{Phys. Rev. X}
  \textbf{\bibinfo{volume}{5}}, \bibinfo{pages}{041041} (\bibinfo{year}{2015}).

\bibitem[{\citenamefont{Takai et~al.}(2016)\citenamefont{Takai, Ido, Misawa,
  Yamaji, and Imada}}]{Takai2016}
\bibinfo{author}{\bibfnamefont{K.}~\bibnamefont{Takai}},
  \bibinfo{author}{\bibfnamefont{K.}~\bibnamefont{Ido}},
  \bibinfo{author}{\bibfnamefont{T.}~\bibnamefont{Misawa}},
  \bibinfo{author}{\bibfnamefont{Y.}~\bibnamefont{Yamaji}}, \bibnamefont{and}
  \bibinfo{author}{\bibfnamefont{M.}~\bibnamefont{Imada}}, \bibinfo{journal}{J.
  Phys. Soc. Jpn.} \textbf{\bibinfo{volume}{85}}, \bibinfo{pages}{034601}
  (\bibinfo{year}{2016}).

\bibitem[{\citenamefont{Yang et~al.}(2010)\citenamefont{Yang, L\"auchli, Mila,
  and Schmidt}}]{PhysRevLett.105.267204}
\bibinfo{author}{\bibfnamefont{H.-Y.} \bibnamefont{Yang}},
  \bibinfo{author}{\bibfnamefont{A.~M.} \bibnamefont{L\"auchli}},
  \bibinfo{author}{\bibfnamefont{F.}~\bibnamefont{Mila}}, \bibnamefont{and}
  \bibinfo{author}{\bibfnamefont{K.~P.} \bibnamefont{Schmidt}},
  \bibinfo{journal}{Phys. Rev. Lett.} \textbf{\bibinfo{volume}{105}},
  \bibinfo{pages}{267204} (\bibinfo{year}{2010}).

\bibitem[{\citenamefont{Kotliar et~al.}(2000)\citenamefont{Kotliar, Lange, and
  Rozenberg}}]{Kotliar2000}
\bibinfo{author}{\bibfnamefont{G.}~\bibnamefont{Kotliar}},
  \bibinfo{author}{\bibfnamefont{E.}~\bibnamefont{Lange}}, \bibnamefont{and}
  \bibinfo{author}{\bibfnamefont{M.~J.} \bibnamefont{Rozenberg}},
  \bibinfo{journal}{Phys. Rev. Lett.} \textbf{\bibinfo{volume}{84}},
  \bibinfo{pages}{5180} (\bibinfo{year}{2000}).

\bibitem[{\citenamefont{Koretsune et~al.}(2007)\citenamefont{Koretsune, Motome,
  and Furusaki}}]{doi:10.1143/JPSJ.76.074719}
\bibinfo{author}{\bibfnamefont{T.}~\bibnamefont{Koretsune}},
  \bibinfo{author}{\bibfnamefont{Y.}~\bibnamefont{Motome}}, \bibnamefont{and}
  \bibinfo{author}{\bibfnamefont{A.}~\bibnamefont{Furusaki}},
  \bibinfo{journal}{J. Phys. Soc. Jpn.} \textbf{\bibinfo{volume}{76}},
  \bibinfo{pages}{074719} (\bibinfo{year}{2007}).

\bibitem[{\citenamefont{Imada}(2005)}]{ImadaMQCP}
\bibinfo{author}{\bibfnamefont{M.}~\bibnamefont{Imada}},
  \bibinfo{journal}{Phys. Rev. B} \textbf{\bibinfo{volume}{72}},
  \bibinfo{pages}{075113} (\bibinfo{year}{2005}).

\bibitem[{\citenamefont{Misawa and Imada}(2007)}]{MisawaImada}
\bibinfo{author}{\bibfnamefont{T.}~\bibnamefont{Misawa}} \bibnamefont{and}
  \bibinfo{author}{\bibfnamefont{M.}~\bibnamefont{Imada}},
  \bibinfo{journal}{Phys. Rev. B} \textbf{\bibinfo{volume}{75}},
  \bibinfo{pages}{115121} (\bibinfo{year}{2007}).

\bibitem[{\citenamefont{Misawa et~al.}(2006)\citenamefont{Misawa, Yamaji, and
  Imada}}]{misawa2006}
\bibinfo{author}{\bibfnamefont{T.}~\bibnamefont{Misawa}},
  \bibinfo{author}{\bibfnamefont{Y.}~\bibnamefont{Yamaji}}, \bibnamefont{and}
  \bibinfo{author}{\bibfnamefont{M.}~\bibnamefont{Imada}}, \bibinfo{journal}{J.
  Phy. Soc. Jpn.} \textbf{\bibinfo{volume}{75}}, \bibinfo{pages}{083705}
  (\bibinfo{year}{2006}).

\bibitem[{\citenamefont{Morita et~al.}(2002)\citenamefont{Morita, Watanabe, and
  Imada}}]{MoritaPIRG}
\bibinfo{author}{\bibfnamefont{H.}~\bibnamefont{Morita}},
  \bibinfo{author}{\bibfnamefont{S.}~\bibnamefont{Watanabe}}, \bibnamefont{and}
  \bibinfo{author}{\bibfnamefont{M.}~\bibnamefont{Imada}},
  \bibinfo{journal}{J.~Phys.~Soc.~Jpn.} \textbf{\bibinfo{volume}{71}},
  \bibinfo{pages}{2109} (\bibinfo{year}{2002}).

\bibitem[{\citenamefont{Kitaev}(2006)}]{AnnalsofPhysics321.2}
\bibinfo{author}{\bibfnamefont{A.}~\bibnamefont{Kitaev}},
  \bibinfo{journal}{Annals Phys.} \textbf{\bibinfo{volume}{321}},
  \bibinfo{pages}{2} (\bibinfo{year}{2006}).

\bibitem[{\citenamefont{Trebst}()}]{trebst2017kitaev}
\bibinfo{author}{\bibfnamefont{S.}~\bibnamefont{Trebst}},
  \eprint{arXiv:1701.07056}.

\bibitem[{\citenamefont{Yamaji et~al.}(2016)\citenamefont{Yamaji, Suzuki,
  Yamada, Suga, Kawashima, and Imada}}]{PhysRevB.93.174425}
\bibinfo{author}{\bibfnamefont{Y.}~\bibnamefont{Yamaji}},
  \bibinfo{author}{\bibfnamefont{T.}~\bibnamefont{Suzuki}},
  \bibinfo{author}{\bibfnamefont{T.}~\bibnamefont{Yamada}},
  \bibinfo{author}{\bibfnamefont{S.-i.} \bibnamefont{Suga}},
  \bibinfo{author}{\bibfnamefont{N.}~\bibnamefont{Kawashima}},
  \bibnamefont{and} \bibinfo{author}{\bibfnamefont{M.}~\bibnamefont{Imada}},
  \bibinfo{journal}{Phys. Rev. B} \textbf{\bibinfo{volume}{93}},
  \bibinfo{pages}{174425} (\bibinfo{year}{2016}).

\bibitem[{\citenamefont{Senthil}(2008)}]{Senthil}
\bibinfo{author}{\bibfnamefont{T.}~\bibnamefont{Senthil}},
  \bibinfo{journal}{Phys. Rev. B} \textbf{\bibinfo{volume}{78}},
  \bibinfo{pages}{035103} (\bibinfo{year}{2008}).

\bibitem[{\citenamefont{Mishmash et~al.}(2015)\citenamefont{Mishmash,
  Gonz\'alez, Melko, Motrunich, and Fisher}}]{MPAFisher}
\bibinfo{author}{\bibfnamefont{R.~V.} \bibnamefont{Mishmash}},
  \bibinfo{author}{\bibfnamefont{I.}~\bibnamefont{Gonz\'alez}},
  \bibinfo{author}{\bibfnamefont{R.~G.} \bibnamefont{Melko}},
  \bibinfo{author}{\bibfnamefont{O.~I.} \bibnamefont{Motrunich}},
  \bibnamefont{and} \bibinfo{author}{\bibfnamefont{M.~P.~A.}
  \bibnamefont{Fisher}}, \bibinfo{journal}{Phys. Rev. B}
  \textbf{\bibinfo{volume}{91}}, \bibinfo{pages}{235140}
  (\bibinfo{year}{2015}).

\bibitem[{\citenamefont{Sengupta et~al.}(2003)\citenamefont{Sengupta, Sandvik,
  and Singh}}]{PhysRevB.68.094423}
\bibinfo{author}{\bibfnamefont{P.}~\bibnamefont{Sengupta}},
  \bibinfo{author}{\bibfnamefont{A.~W.} \bibnamefont{Sandvik}},
  \bibnamefont{and} \bibinfo{author}{\bibfnamefont{R.~R.~P.}
  \bibnamefont{Singh}}, \bibinfo{journal}{Phys. Rev. B}
  \textbf{\bibinfo{volume}{68}}, \bibinfo{pages}{094423}
  (\bibinfo{year}{2003}).

\bibitem[{\citenamefont{Watanabe et~al.}(2014)\citenamefont{Watanabe, Kawamura,
  Nakano, and Sakai}}]{doi:10.7566/JPSJ.83.034714}
\bibinfo{author}{\bibfnamefont{K.}~\bibnamefont{Watanabe}},
  \bibinfo{author}{\bibfnamefont{H.}~\bibnamefont{Kawamura}},
  \bibinfo{author}{\bibfnamefont{H.}~\bibnamefont{Nakano}}, \bibnamefont{and}
  \bibinfo{author}{\bibfnamefont{T.}~\bibnamefont{Sakai}}, \bibinfo{journal}{J.
  Phys. Soc. Jpn.} \textbf{\bibinfo{volume}{83}}, \bibinfo{pages}{034714}
  (\bibinfo{year}{2014}).

\bibitem[{\citenamefont{Kawamura et~al.}(2017)\citenamefont{Kawamura, Yoshimi,
  Misawa, Yamaji, Todo, and Kawashima}}]{hphi}
\bibinfo{author}{\bibfnamefont{M.}~\bibnamefont{Kawamura}},
  \bibinfo{author}{\bibfnamefont{K.}~\bibnamefont{Yoshimi}},
  \bibinfo{author}{\bibfnamefont{T.}~\bibnamefont{Misawa}},
  \bibinfo{author}{\bibfnamefont{Y.}~\bibnamefont{Yamaji}},
  \bibinfo{author}{\bibfnamefont{S.}~\bibnamefont{Todo}}, \bibnamefont{and}
  \bibinfo{author}{\bibfnamefont{N.}~\bibnamefont{Kawashima}},
  \bibinfo{journal}{Comput. Phys. Commun.} \textbf{\bibinfo{volume}{217}},
  \bibinfo{pages}{180} (\bibinfo{year}{2017}).

\bibitem[{hph({\natexlab{a}})}]{hphi_ma}
\bibinfo{note}{~http://ma.cms-initiative.jp/en/application-list/hphi}.

\bibitem[{hph({\natexlab{b}})}]{hphi_git}
\bibinfo{note}{~https://github.com/QLMS/HPhi}.
\end{thebibliography}

\end{document}